\documentclass[reprint,pra,nofootinbib]{revtex4-1}

 \usepackage{amssymb}
 \usepackage{dsfont}
 \usepackage{mathdots}
 \usepackage{amsmath}
\usepackage{graphicx}
\usepackage{dcolumn}
\usepackage{bm}
\usepackage{color,soul}
\usepackage{relsize}

\begin{document}
\title{State transfer based on classical nonseparability}
\author{\mbox{Seyed Mohammad Hashemi Rafsanjani$^1$*}}
\author{\mbox{Mohammad Mirhosseini$^1$*}}
\author{Omar S. Maga\~{n}a-Loaiza$^1$}
\author{ Robert W. Boyd$^{1,2}$}
\affiliation{$^1$ Institute of Optics, University of Rochester, Rochester, New York 14627\\ $^2$Department of Physics, University of Ottawa, Ottawa, ON K1N6N5 Canada}
\email{hashemi@pas.rochester.edu\\mirhosse@optics.rochester.edu\\ These authors, SMHR and MM, have contributed equally to this work.}



\date{\today}

\begin{abstract} 
We present a state transfer protocol that is mathematically equivalent to quantum teleportation, but uses classical nonseparability instead of quantum entanglement. In our implementation we take advantage of nonseparability among three parties: orbital angular momentum (OAM), polarization, and the radial degrees of freedom of a beam of light. We demonstrate the transfer of arbitrary OAM states, in the subspace spanned by any two OAM states, to the polarization of the same beam.
\end{abstract}

\pacs{..........}

\maketitle

 \newcommand{\beq}{\begin{equation}}
 \newcommand{\eeq}{\end{equation}}
 \newcommand{\bel}{\begin{align*}}
 \newcommand{\tamam}{\end{align*}}
 \newcommand{\dg}[1]{#1^{\dagger}}
 \newcommand{\reci}[1]{\frac{1}{#1}}
 \newcommand{\ket}[1]{|#1\rangle}
 \newcommand{\nim}{\frac{1}{2}}
 \newcommand{\om}{\omega}
 \newcommand{\te}{\theta}
 \newcommand{\la}{\lambda}
 \newcommand{\beqa}{\begin{eqnarray}}             
 \newcommand{\eeqa}{\end{eqnarray}}               
 \newcommand{\nn}{\nonumber}                      
 \newcommand{\bra}[1]{\langle#1\vert}                 
 \newcommand{\ipr}[2]{\left\langle#1|#2\right\rangle}
  \newcommand{\up}{\uparrow}
  \newcommand{\down}{\downarrow}
  \newcommand{\dn}{\downarrow}         
   \newcommand{\cket}[1]{|#1)}
    \newcommand{\cbra}[1]{(#1\vert}       
    \newcommand{\blu}[1]{\textcolor{blue}{#1}} 
        \newcommand{\red}[1]{\textcolor{red}{#1}} 
        
\section{Introduction}
Entanglement in quantum systems leads to many of the surprising consequences of the quantum mechanical description of nature. For many decades the aim of physicists has been to realize and confirm such phenomena in experimental settings. Such efforts came to fruition in a series of seminal observations that validated quantum mechanics and contradicted some classical alternatives such as local hidden variable theories \cite{PhysRevLett.28.938,PhysRevLett.47.460,PhysRevLett.49.1804,PhysRevLett.81.5039}.
        
Although entanglement is often thought as an exclusively-quantum phenomenon, the mathematical structure behind it that quantifies the degree of nonseparablity can be applied to any two vector spaces. In fact, in his seminal 1935 paper Schr\"odinger \cite{PSP:1737068,*PSP:2027212} pointed out that the mathematics that he utilized was already known by mathematicians \cite{schmidt}. This mathematical structure when applied to describe the non-separability of different degrees of freedom has come to be known as \textit{classical entanglement} \cite{spreeuw1998classical,kagalwala2013bell,toppel2014classical}. 
Although the analogy between quantum entanglement and the classical entanglement stops when non-locality comes into picture, the identification of their similarity has proven to be helpful in developing a new perspective in determining the degree of polarization of a beam of light \cite{spreeuw1998classical,PhysRevLett.104.023901,qian2011entanglement,kagalwala2013bell}. Furthermore, the violation of Bell inequalities between different degrees of freedom (DoF) of a beam of light has been the subject of several notes recently \cite{PhysRevA.82.033833, PhysRevA.82.022115, PhysRevA.90.052326, PhysRevA.90.053842, kagalwala2013bell, qian2014violation}.

The above analogy between quantum entanglement and its classical analog can be extended to multipartite systems \cite{PhysRevA.63.062302, aiello2014classical}. In the original proposal by \citet{PhysRevA.63.062302}, a beam is divided into several beams and by controlling the amplitude and phase of each portion of the beam one can mimic, a GHZ-state, some quantum gates, and teleportation. One of the most well-known consequence of entanglement that involves more than two parties is the phenomenon of teleportation, that was first proposed by \citet{PhysRevLett.70.1895} and was realized with quantum entanglement by \citet{Quantumteleportation}. Since then teleportation has been proposed and realized in different systems \cite{PhysRevLett.80.1121,*PhysRevLett.86.1370,*riebe2004deterministic,*barrett2004deterministic,*olmschenk2009quantum,*PhysRevA.83.060301,*erhard2014real}. 

Teleportation allows us to transfer the state of one party to a non-local party via a projection on a Bell state. If the non-local parties are replaced by different DoFs of a beam we end up with a procedure to transfer state of one degree of freedom to another through a Bell-like projection. Although this may not be quite as intriguing a phenomenon as teleportation, the capability to transfer an arbitrary, and a priori unknown, state from one degree of freedom to another is a non-trivial, and desirable task \cite{PhysRevLett.96.163905,bozinovic2013terabit,shen2015integrated}. In the following we report on the realization of this phenomenon, i.e. state transfer between two DoFs in an approach that mimics teleportation. In our implementation we \textit{transfer} an arbitrary state of any two orbital angular momentum (OAM) modes of a laser beam onto the polarization of the same laser beam. Although the nonseparability between different degrees of freedom of a laser beam has come to be referred to as \emph{classical entanglement}, one can argue that the term entanglement should be reserved for the cases that involves inherently quantum mechanical systems that cannot be described classically. Thus we reserve the term entanglement for quantum entanglement and refer to its classical analog as classical nonseparability. The term classical nonseparability simultaneously captures two different aspects: First that the experiment that we are upon taking can be described without invoking of quantum mechanics, and is hence \emph{classical}, and secondly that such nonseparability in optical beams bares merely a mathematical similarity to its quantum counterpart. 

The manuscript is organized as follows. In the next section we explain the  procedure of the state transfer protocol as an analog of teleportation when the non-local parties are replaced by DoFs. In section \ref{iii} we explain the details of our experiment and present a discussion on our results and their implication. Concluding considerations are presented in section \ref{conclusion}.

\section{State transfer as an analog of Teleportation}
The original proposal by \citet{PhysRevLett.70.1895} takes advantage of three parties. One we call Alice, the other Bob, and a third party that we name Charlie. Initially Alice's state is separable from the other two parties and Charlie and Bob share a joint maximally entangled state. One then performs a projective measurement on a joint Bell state of Alice and Charlie. When we post-select on those measurements that have led to a specific Bell state, Bob's state will be the same as the initial Alice's state.

\begin{figure}[tbp]
\includegraphics[width=\columnwidth]{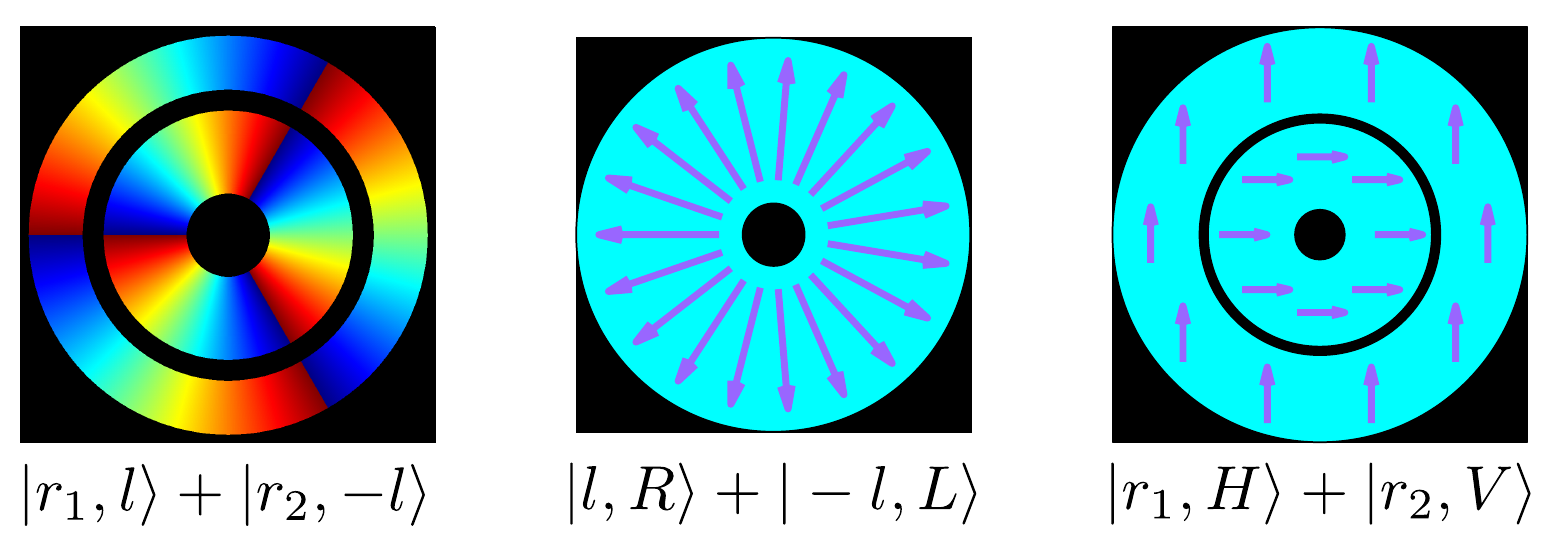}
\caption{Three examples of classical nonseparability. The left panel represents a beam for which the radial and orbital angular momentum ($l=3$) degrees of freedom are nonseparable. The middle panel shows a beam for which the orbital angular momentum ($l=1$) and polarization are nonseparable. The panel on right represents a beam for which radial and polarization degrees of freedom are nonseparable.}
\label{examples}
\end{figure}

To realize our protocol for coherence transfer we replace the three parties with three degrees of freedom of a single optical beam.\,In our implementation, the three degrees of freedom are the the radial degree of freedom, polarization, and orbital angular momentum, that play the roles of Charlie, Bob, and Alice respectively. In Fig.\,\ref{examples} we have presented three examples of nonseparability between different degrees of freedom of a spatial profile of a beam of light. These can be considered analogs of Bell states between different DoFs that we deal with in this paper. In principle any three DoFs can be used to realize such state transformation protocol as long as one can perform arbitrary joint measurements on these observables/quantities. 

\begin{figure}[tp]
\includegraphics[width=\columnwidth]{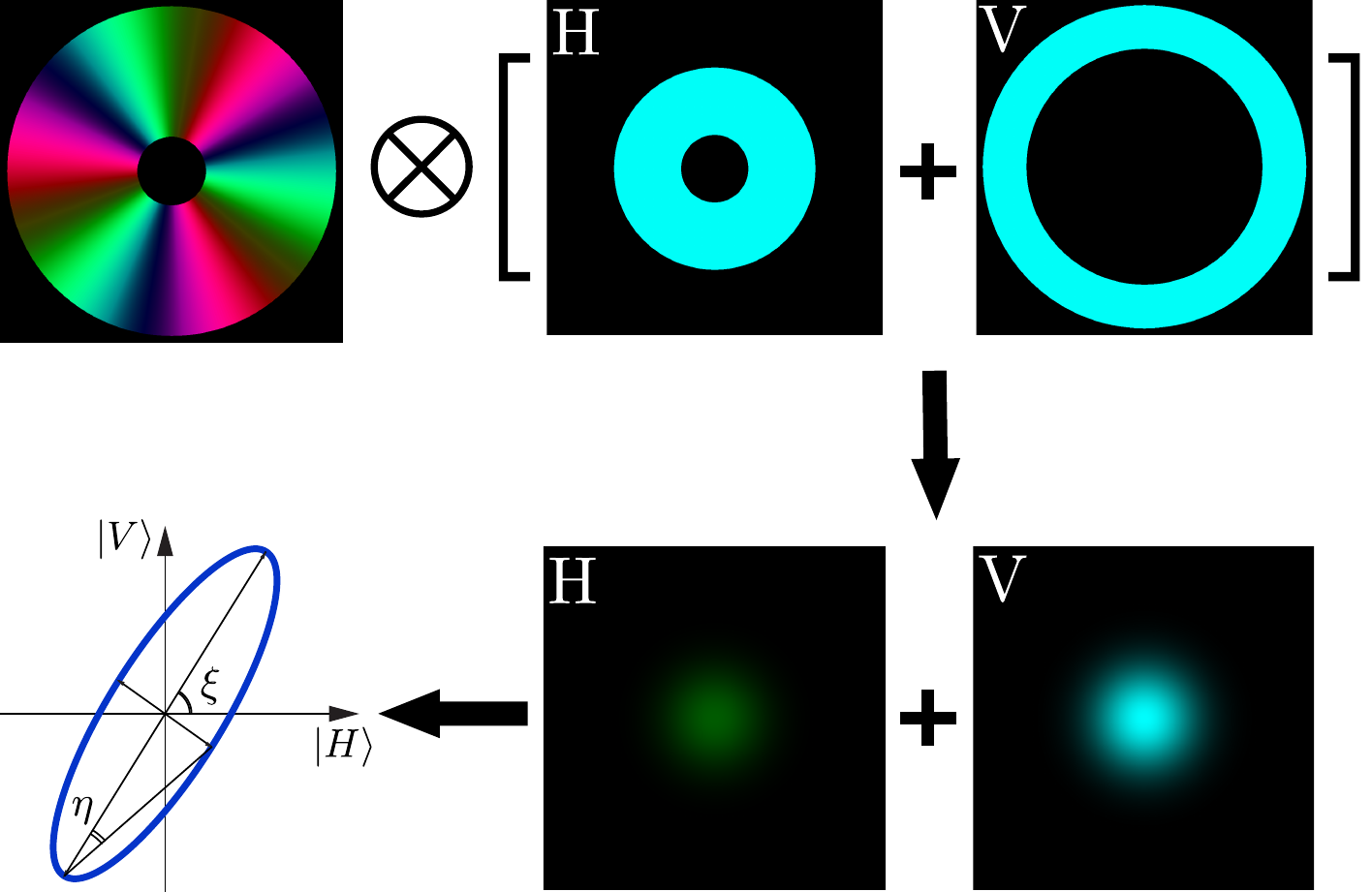}
\caption{Transfer of an arbitrary OAM state to a polarization.  The intensity is encoded in the brightness, the phase in the color. The OAM information can be read from the phase profile. For simplicity we have depicted a pure OAM superposition of $\ket{\pm 3}$. (top) The initial beam: OAM is separable from the other two DoFs and the polarization and radial DoFs are maximally nonseparable. (bottom, right) After projection on a maximally nonseparable state of OAM and radial, we end up with a polarization state that carries the same information as the initial OAM state. The intensity of each of polarization gives information about the amplitude of each of the two OAM components and the phase between the two polarizations, $H$ and $V$, is the same as the phase between the two OAM components. (bottom, left) The two angles $\xi$, and $\eta$ are half of the spherical angles on the Poincare sphere, respectively.}
\label{protocol}
\end{figure}

In our realization we first produce a beam whose polarization and radial DoFs are nonseparable and both are separable from the OAM:
\begin{align}
\biggl[\gamma \ket{l}+\bar{\gamma}\ket{-l}\biggr]\otimes \biggl[\ket{r_{1},H}+\ket{r_{2},V}\biggl].
\label{initial_state}
\end{align}
\noindent $\ket{H},\ket{V}$ denote the horizontal and vertical polarizations. We emphasize that although we adopt the ket-bra notation that is associated with quantum mechanics, the description of our experiment requires no invoking of quantum mechanics and we adopt this notation to emphasize the linear algebraic nature of different degrees of freedom. The polarization of an optical field arises from the vectorial nature of electromagnetic field and techniques for its manipulation are easy to implement. $\ket{l}$ denotes an OAM mode, which is defined via the helical phase structure $e^{i\ell \phi}$. OAM modes naturally arise as paraxial solutions to the Maxwell equations in cylindrical coordinates, and hence can be completely understood using wave optics \cite{PhysRevA.45.8185,PhysRevLett.73.1239}. Nevertheless, the OAM modes can be useful in quantum optics too \cite{leach2004laser,*molina2007twisted,*PhysRevLett.105.153601,*mirhosseini2013efficient,PhysRevLett.96.163905}. Finally $\ket{r_1},\ket{r_2}$ denote two radial modes, defined as two concentric, mutually exclusive, annular regions with a uniform intensity pattern. Note that $\ket{r_1},\ket{r_2}$ are orthogonal since there is no overlap between their corresponding spatial extents. Radial modes have also been the subject of a few recent investigations for their potential applicability in quantum communication \cite{PhysRevA.83.033816,*PhysRevLett.108.173604, *karimi2012radial, *krenn2014generation, *PhysRevA.89.013829}.  The transverse profile of a beam represented by Eq.\,(\ref{initial_state}) is depicted in Fig.\,\ref{protocol} (top). The dependence of the phase on the azimuthal angle is identical for both radial components since their OAM contents are the same. 
  
Now that we have prepared the state in Eq.\,(\ref{initial_state}), the next step in the protocol is to implement a projective measurement onto one of the Bell states of Charlie(radial)-Alice(OAM):
\begin{align} 
    &\bra{\Phi^\pm}=\bra{r_1,l}\pm \bra{r_2,-l}, \\[0.15cm] \nn
    &\bra{\Psi^\pm}=\bra{r_1,-l}\pm \bra{r_2,l}. 
 \end{align}
Depending on our choice, the state of Bob (polarization) will be either $\gamma \ket{H}\pm \bar{\gamma}\ket{V}$ for projection onto $\ket{\Phi^\pm}$, and $\gamma \ket{V}\pm \bar{\gamma}\ket{H}$ for projection onto $\ket{\Psi^\pm}$ respectively. We choose to project onto $\ket{\Phi^+}$. In our projection we use a pinhole in the far field (setup below); thus after the projection the light emerging from the pinhole  represents a single spatial mode that carries no orbital angular momentum. The transverse profile of such a beam is depicted in Fig.\,\ref{protocol} (bottom, right). The polarization of this beam is completely separable from the radial and OAM DoFs and the emerging beam's polarization state reads:
 \begin{align}
\gamma \ket{H}+\bar{\gamma}\ket{V}.
\end{align}
We note that the final Bob (polarization) state carries the same information as the initial Alice (OAM) state in Eq.\,(\ref{initial_state}). This result is independent of the the choice of the initial state. Although our derivation has assumed that Alice's initial state to be a pure state, the derivation can be easily generalized to accommodate mixed states \cite{PhysRevLett.70.1895}.

\section{Experimental implementation}\label{iii}
\begin{figure*}[htbp]
\includegraphics[width=0.7\paperwidth]{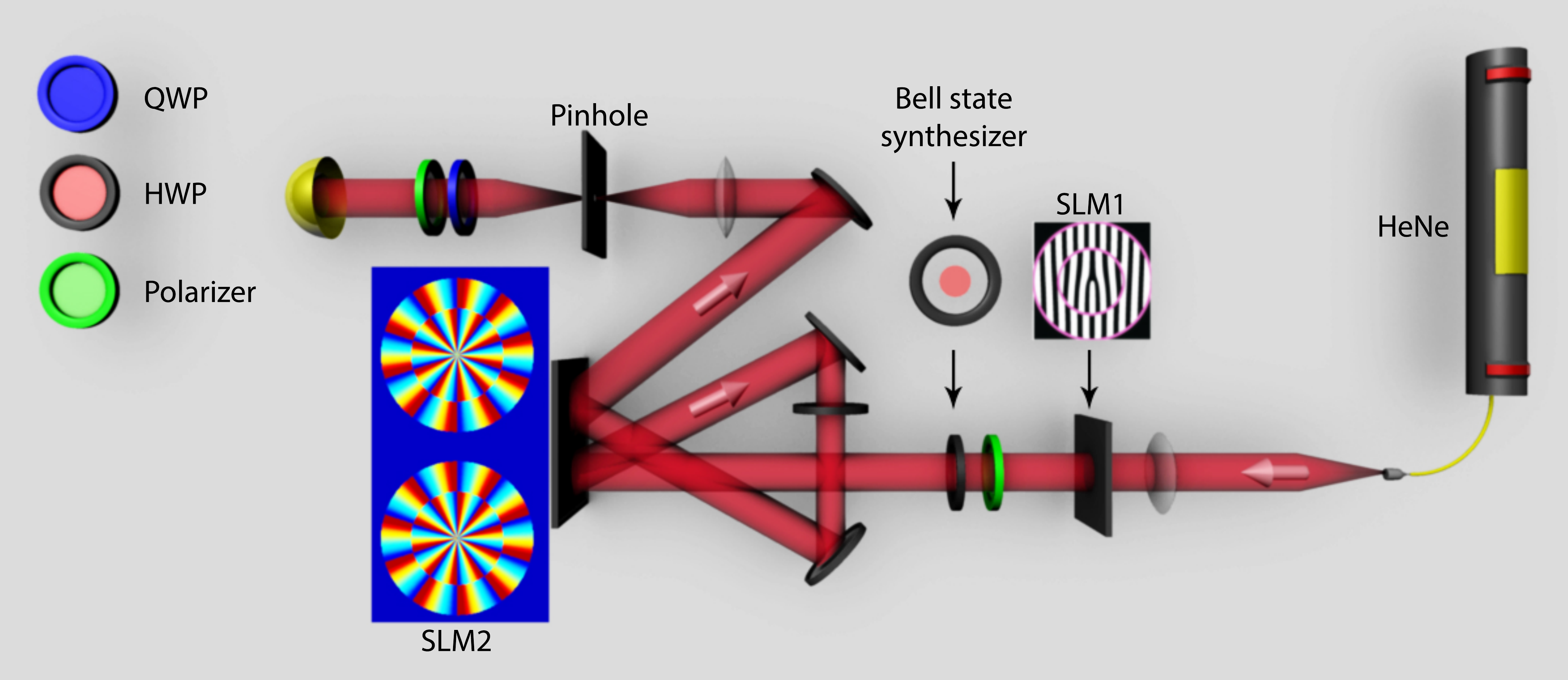}
\caption{The setup to implement the state transfer from OAM to the polarization DoF; the beam emerging from a single mode fiber is collimated and shined onto a spatial light modulator (SLM1). The beam emerging from the SLM1 has two rings with identical azimuthal phase profiles (OAM states). Using a polarizer and a small half wave plate (Bell-state synthesizer), orthogonal polarizations are introduced onto the two rings. To implement projection onto a radial-OAM maximally nonseparable state, we divide the surface of SLM2 to two parts. The beam is first shined onto one half of SLM2 where we have impressed a phase screen that has two rings with opposite OAM values. Then we use a half wave plate to rotate the polarization of the beam 90 degrees and then shine the light on the second half of SLM2. This combination allows us to perform polarization insensitive projections onto the radial-OAM maximally nonseparable state. To complete the projection we use a pinhole to separate the projected light and then use different combinations of a quarter wave plate and a polarizer to measure the Stokes parameters.}
\label{schematic}
\end{figure*}

A schematic representation of the setup is given in Fig.\,\ref{schematic}. Our source of light is a cw He-Ne laser that emits at the wavelength of 633 nm. In order to produce the state prepared in Eq.\,(\ref{initial_state}) we first use a hologram to produce a coherent beam of two rings. The phase profiles of both rings are identical and match the OAM state that is to be teleported. In principle, one can choose to use any two orthogonal OAM state. In our realization we used the two OAM states $\{\ket{10}, \ket{-10}\}$. This choice minimizes the cross talk between the two states that often results from imperfect experimental realization of OAM projections. We use a telescope after the SLM1 to increase the diameter of the initial beam to about 2 cm.

The laser beam is then passed through a polarizer and then a half wave plate (HWP) whose aperture only covers the inner portion of the beam. The HWP is set to $45^\circ$ in order to rotate the polarization of the inner disk to the orthogonal polarization. We name this combination a Bell-state synthesizer. After rotating the polarization of the inner beam using a small half-wave plate, we measure the power contained in each of the two rings.  We then match their power by adjusting the radii of the two rings using SLM1. The last HWP induces a phase difference between the two rings that can be cancelled by the spatial light modulator used for shaping the laser beam [Fig \ref{schematic}, SLM1]. 
The beam emerging from the last HWP can be set to possess an arbitrary state of OAM, along with a polarization structure that is maximally nonseparable from the radial DoF. As a result, the field can be formally described by Eq. (\ref{initial_state}). At this stage, we need to project onto a joint maximally nonseparable state of OAM and radial degrees of freedom in order to realize the state conversion. The SLM allows for performing a projection onto the OAM state of $l=-10$ in the inner disk and a simultaneous projection over $l=10$ for the outer annular ring. We use a phase-only liquid crystal SLM to shape the wavefront of the horizontal polarization component of the beam. To achieve a polarization-insensitive projection, we use the SLM in a double-pass geometry, with a HWP in between the two reflections for rotating the polarization of the beam by $90^\circ$.  In both reflections SLM1 is imaged on SLM2 (the imaging optics is not shown in the figure). We use a lens with a focal length of 30 cm after the SLM to focus the beam onto a pinhole with a diameter of 5 microns. The beam emerging from the pinhole is approximately a single spatial mode with a polarization state that is related to the initial OAM state of Alice. A combination of a polarizer, quarter wave plate, and detector are used to measure the stokes parameters and subsequently characterize the polarization state.
\begin{figure}[tbp]
\includegraphics[width=\columnwidth]{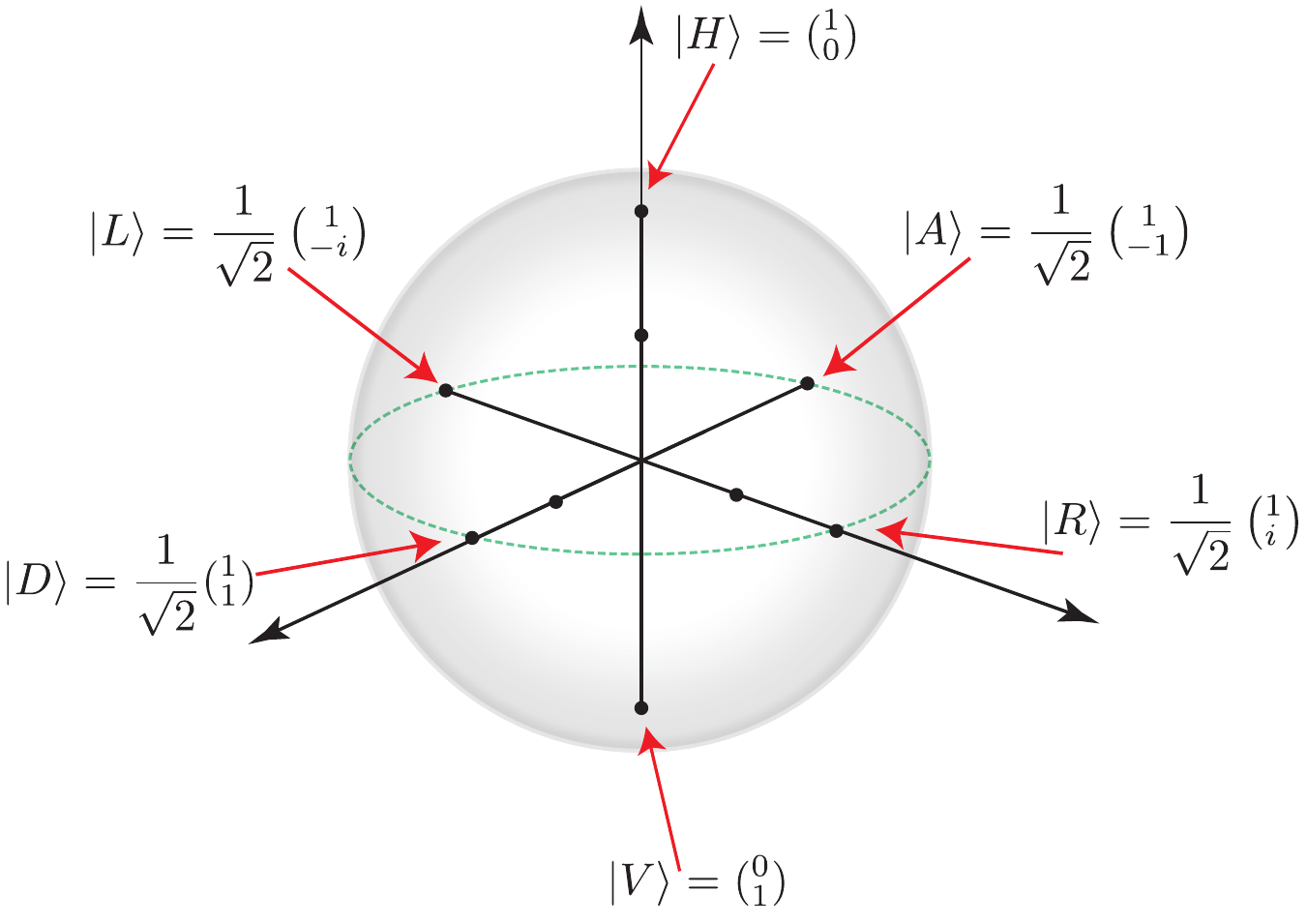}
\vspace{0.5cm}
\begin{tabular}{ |c | c | c |}
 \hline           
 &  & Fidelity of \\
Alice's state  &  Ideal Bob's state &  transfer \\
\hline \hline
$\ket{h_l}$ & $\ket{H}$ & $98.8\pm 0.4$ \\
$\ket{v_l}$ & $\ket{V}$ & $98.8\pm 0.2$ \\
$\ket{d_l}$ & $\ket{D}$ & $98.9\pm 0.5$ \\
$\ket{a_l}$ & $\ket{A}$ & $99.3\pm 0.3$~\\
$\ket{r_l}$ & $\ket{R}$ & $99.1\pm 0.4$\\
$\ket{l_l}$& $\ket{L}$ & $99.2\pm 0.3$ \\
$3\ket{h_l}\bra{h_l}+\ket{v_l}\bra{v_l}$& ~$3\ket{H}\bra{H}+\ket{V}\bra{V}$ & $99.5\pm 0.4$ \\
$3\ket{d_l}\bra{d_l}+\ket{a_l}\bra{a_l}$& ~$3\ket{D}\bra{D}+\ket{A}\bra{A}$ & $99.3\pm 0.3$ \\
$3\ket{r_l}\bra{r_l}+\ket{l_l}\bra{l_l}$ & ~$3\ket{R}\bra{R}+\ket{L}\bra{L}$ & $99.2\pm 0.2$ \\
\hline
\end{tabular}
\caption{Fidelity of state transfer for different states: All OAM states are in the subspace spanned by $l=\pm 10$. The three last states are mixed states. For brevity of notation, we have shown the unnormalized states.}
\label{fidelities} 
\end{figure}

We test our protocol by first transferring pure states of OAM. This has been done by converting the polarization Stokes parameters into a two-dimensional Jones vector and then finding the degree of similarity between the initial (OAM) state owned by Alice and the final (polarization) state detected by Bob. In Fig.\,\ref{fidelities} we report the fidelities between different initial OAM states and the polarization state that was measured at the end.  

The initial OAM states are chosen to be along the three primary axes of the Bloch sphere for a two-dimensional sub-space of $\{\ket{10},\ket{-10}\}$. Namely, we have transferred the states $\{\ket{10},\ket{-10}, \ket{10}+\ket{-10},\ket{10}-\ket{-10},\ket{10}+i\ket{-10}, \ket{10}-i\ket{-10} \}$. The ideal converted states are then supposed to be the following polarization state respectively: $\{\ket{H},\ket{V}, \ket{H}+\ket{V},\ket{H}-\ket{V},\ket{H}+i\ket{V}, \ket{H}-i\ket{V} \}$. The following equation provides the corresponding mapping between OAM and polarization states:
\begin{align}\nn
\ket{h_l}&=\ket{10}&\hspace{-1.5cm}\longmapsto\ket{H},\\ \nn
\ket{v_l}&=\ket{-10}&\hspace{-1.5cm}\longmapsto\ket{V},\\ \nn
\ket{d_l}&=(\ket{10}+\ket{-10})/\sqrt{2}&\hspace{-1.5cm}\longmapsto\ket{D},\\ \nn
\ket{a_l}&=(\ket{10}-\ket{-10})/\sqrt{2}&\hspace{-1.5cm}\longmapsto \ket{A},\\       \nn
\ket{r_l}&=(\ket{10}+i\ket{-10})/\sqrt{2}&\hspace{-1.5cm}\longmapsto\ket{R},\\  
\ket{l_l}&=(\ket{10}-i\ket{-10})/\sqrt{2}&\hspace{-1.5cm}\longmapsto\ket{L}.
\end{align}

Note that although the quantum density matrix is by definition a semi-positive definite matrix, the results of state tomography for a pure state often turns out to have negative eigenvalues \cite{blume2010optimal}. This is primarily due to imperfect projective measurements and the noise in the experiment. We have used the maximum likelihood recovery algorithm to find a positive state that is the most probable given the data from the measurement. The average fidelity of transferred states with their corresponding initial states is approximately $99\%$, demonstrating a notably good agreement with the theoretical predictions.

From a practical point of view pure states are an idealization; irrespective of how carefully a state is prepared, noise will inevitably render a pure state mixed. It is then significant if an implementation can also accommodate mixed states. Additionally, pure states are only a restricted set of physical states in the Hilbert space. The vast majority of states are mixed states \cite{blume2010optimal}. Since we always project onto the same Bell OAM-radial state, our implementation allows us to also transfer the mixed states. For demonstration we have also transferred three typical mixed states. To produce mixed states we randomly switch the hologram on SLM1 and use a long (10 minutes) integration time using a power-meter. We randomly switch between two holograms on SLM1 such that $75\%$ of the time we prepare one pure OAM state and $25\%$ we prepare another pure state. 
The states are chosen to be $0.75\ket{h}\bra{h}+0.25\ket{v}\bra{v}$, $0.75\ket{d}\bra{d}+0.25\ket{a}\bra{a}$, and $0.75\ket{r}\bra{r}+0.25\ket{l}\bra{l}$. These OAM states are ideally teleported to the polarization states $0.75\ket{H}\bra{H}+0.25\ket{V}\bra{V}$, $0.75\ket{D}\bra{D}+0.25\ket{A}\bra{A}$, and $0.75\ket{R}\bra{R}+0.25\ket{L}\bra{L}$, respectively. In Fig.\,\ref{fidelities} we have reported the fidelities between the polarization states from the experiment with the ones from theory. Note that the fidelity between two mixed states is defined as $\rho, \sigma$ is $F=\rm{tr} \sqrt{\rho^{1/2} \sigma \rho^{1/2}}$. The average fidelities for the three representative mixed states are found to be 99.33\%, which confirms the accurate operation of our experimental realization.

Considering that the formalism of classical nonseparability applies to any three degrees of freedom, we anticipate that this machinery can be utilized to transfer the state of other DoFs to another, provided the technical complication in performing the appropriate rotations and projections on other DoFs can be met. This is a non-trivial problem. For example, the radial degree of freedom that was utilized here as an ancilla has attracted a lot of interest recently \cite{PhysRevA.83.033816,*PhysRevLett.108.173604,*karimi2012radial,*krenn2014generation, *PhysRevA.89.013829}. Nonetheless the problem of efficient projection on arbitrary Laguerre-Gaussian modes remain a challenge.

Our specific example provides the capability to map an arbitrary state of any two OAM modes to a polarization state in a one-to-one fashion. Given the important role that transferring states between different degrees of freedom plays in recent experiments in quantum information science  \cite{PhysRevLett.103.013601, nagali2009optimal}, we believe that our specific example has the potential for a wide range of applications. It should be noted that there are also other approaches that can be used to transfer OAM state onto the polarization of the beam \cite{d2012deterministic,karimi2010polarization}.

\section{conclusion}\label{conclusion}
While entanglement is an essential part of the quantum paradigm, the mathematical idea behind it, i.e. nonseparability, may manifest itself in systems whose description does not requiring the invoking of quantum theory. Careful examination of such analogy may help us develop a modern perspective toward old concepts \cite{PhysRevLett.104.023901,qian2011entanglement} as well as develop techniques whose simplicity has not been appreciated, and/or whose applicability have not been fully exhausted yet.

In this manuscript we took inspiration from the phenomenon of teleportation to develop a state transfer protocol that is mathematically equivalent to teleportation, but uses classical nonseparability instead of entanglement. Initially the OAM modes are separable from the radial and polarization of degree of freedom, and polarization and the radial degree of freedom share a maximally nonseparable state. Then we implement a Bell state projection on 
OAM-radial degree of freedom. The polarization of the beam emerging from this projection carries the same information as the initial OAM states. Our protocol allows one to transfer an arbitrary, and a priori unknown state of any two OAM mdoes of a beam of light to its polarization state. \\

\section{acknowledgement}
We acknowledge helpful discussions with A. Aiello and J. H. Eberly. RWB acknowledges funding from the Canada Excellence Research Chairs program.\, OSML acknowledges support from the Consejo Nacional de Ciencia y Tecnolog\'ia (CONACyT), the Secretar\'ia de Educaci\'on P\'ublica (SEP), and the Gobierno de M\'exico.
\bibliographystyle{apsrev4-1}
\bibliography{mybib}

\end{document}